 \documentclass[12pt,preprint]{aastex}

\usepackage{graphicx}				         
\usepackage{amsmath}                                       
\usepackage{enumitem}			          	
\usepackage{amssymb}
\usepackage{pdfpages}
\usepackage{hyperref}

\newcommand{\Lya}     {Ly$\alpha$}    

\newcommand {\HI}        {\ion{H}{1}}   

\newcommand {\HeI}     {\ion{He}{1}}   

\newcommand {\OI}      {\ion{O}{1}}   

\newcommand {\NI}       {\ion{N}{1}}

\newcommand {\SII}       {\ion{S}{2}}

\newcommand {\kms}    {km~s$^{-1}$}

\newcommand {\etal}   {et~al.}

\begin{document}

\title{Epsilon Canis Majoris:  The Brightest EUV Source with \\
Surprisingly  Low Interstellar Absorption} 

\author {J. Michael Shull$^{1,2}$, Rachel M. Curran$^2$, and Michael W. Topping$^3$} 

\affil{(1) Department of Astrophysical \& Planetary Sciences, \\ 
          University of Colorado; Boulder CO 80309, USA; \\
        (2) Department of Physics and Astronomy, \\
           University of North Carolina, Chapel Hill NC 27599, USA;   \\ 
        (3) Department of Astronomy, Steward Observatory, \\
             University of Arizona, Tucson AZ  85721, USA }

\email{michael.shull@colorado.edu, rcurran@unc.edu, \\ michaeltopping@arizona.edu} 


\begin{abstract}

The B2 star $\epsilon$~CMa, at parallax distance $d = 124\pm2$~pc, dominates the 
\HI\ photoionization of the local interstellar cloud (LIC).  At its closer parallax distance 
compared to previous estimates, $\epsilon$~CMa has a 0.9 mag fainter absolute 
magnitude $M_V =-3.97\pm0.04$.  We combine measurements of distance with the
integrated flux $f = (41.5\pm3.3) \times 10^{-6}~{\rm erg~cm}^{-2}~{\rm s}^{-1}$ and 
angular diameter $\theta_d = 0.80\pm0.05$~mas to produce a consistent set of stellar 
parameters:  radius $R = 10.7\pm0.7~R_{\odot}$, mass $M = 13.1\pm2.3~M_{\odot}$, 
gravity $\log g = 3.50\pm0.05$, effective temperature $T_{\rm eff} \approx 21,000$~K, 
and luminosity $L \approx 20,000~L_{\odot}$.   These parameters place $\epsilon$~CMa 
outside the $\beta$ Cephei  instability strip, consistent with its observed lack of pulsations.  
The observed EUV spectrum yields a hydrogen photoionization rate 
$\Gamma_{\rm HI} \approx 10^{-15}$~s$^{-1}$ (at Earth). The total flux decrement factor 
at the Lyman limit ($\Delta_{\rm LL} = 5000\pm500$) is a combination of attenuation 
in the stellar atmosphere ($\Delta_{\rm star} = 110\pm10$) and interstellar medium
($\Delta_{\rm ISM} = 45\pm5$) with optical depth $\tau_{\rm LL} = 3.8\pm0.1$.
After correcting for interstellar \HI\ column density 
$N_{\rm HI} = (6\pm1)\times10^{17}~{\rm cm}^{-2}$, we find a stellar LyC photon flux 
$\Phi_{\rm LyC} \approx 3000~{\rm cm}^{-2}~{\rm s}^{-1}$ and ionizing luminosity 
$Q_{\rm LyC} = 10^{45.7\pm0.3}$~photons~s$^{-1}$.  The photoionization rate 
$\Gamma_{\rm H} \approx$ (1--2)$\times 10^{-14}~{\rm s}^{-1}$ at the cloud surface
produces an ionization fraction (30--40\%) for total hydrogen density
$n_{\rm H} = 0.2$ cm$^{-3}$. With its $27.3\pm0.4$~\kms\ heliocentric radial velocity 
and small proper motion, $\epsilon$~CMa  passed within $9.3\pm0.5$ pc of the Sun 
4.4 Myr ago, with a 180 times higher photoionization rate.  

\end{abstract}

\section{Introduction to Properties of $\epsilon$~CMa}  

The B-type giant star Epsilon~Canis Majoris ($\epsilon$~CMa), also known as HD~52089 and 
Adhara, is the brightest source of extreme ultraviolet (EUV) radiation in the sky (J.\ Dupuis \etal\ 1995; 
J.\ Vallerga \& B.\ Welsh 1995) as observed by the Extreme Ultraviolet Explorer (EUVE).  Because of 
the low-density interstellar medium (ISM) or ``local interstellar tunnel" along its sight line 
(Gry \etal\ 1985;  B.\ Welsh 1991; J.\ Cassinelli \etal\ 1995; J.\ Vallerga 1998; J.\ Linsky \etal\ 2019), 
this star dominates the local \HI\ photoionization (J.\ Vallerga 1998), with a rate of 
$\Gamma_{\rm HI} \approx 1.0\times10^{-15}$~s$^{-1}$ as viewed from Earth.  Its spectrum, shown 
in Figure 1, has been measured by EUVE at $\lambda \leq 730$~\AA, and more recently with the 
Colorado-DEUCE rocket at 700--1150~\AA\ (N.\ Erickson \etal\ 2021).  The ionizing continuum flux 
outside the local interstellar clouds is considerably higher, after correcting for photoelectric absorption 
by \HI\ ($\lambda \leq 912$~\AA) and \HeI\ ($\lambda \leq 504$~\AA).  

In a classic survey of southern B-type stars, J.\ Lesh (1972) listed the spectral type (SpT) of
$\epsilon$~CMa as B2~II, with apparent and absolute visual magnitudes $(m_V = 1.50$, 
$M_V = -4.9$) and a spectrophotometric distance of 187~pc.  In later papers, the distance to 
$\epsilon$~CMa was quoted variously as 179~pc (R.\ Bohlin 1975) and 188~pc 
(e.g., T.\ Snow \& D.\ Morton 1976;  B.\ Savage \etal\ 1977; R.\ Bohlin \etal\ 1978;  
J.\ Cassinelli \etal\ 1995).  However, the {\it Hipparcos} parallax measurement of 
$8.05\pm0.14$~mas (F.\ van Leeuwen 2007) provided a shorter distance $d = 124\pm2$~pc, 
with distance modulus ($m_V$--$M_V) = 5.467 \pm0.035$ smaller by 0.90 mag than for 188~pc.  
The observed visual magnitude $m_V = 1.50$ implies an absolute magnitude $M_V = -3.97 \pm 0.04$,
also 0.9~mag fainter than the value of $-4.9$ given in J.\ Lesh (1972) for B2~II type.  The luminosity 
gap is even larger for the recent classification as B1.5~II (L.\ Fossati \etal\ 2015) with $M_V = -5.1$.  
Unless the parallax measurements are wrong, the star is underluminous for its morphologically 
classified spectral type.

In this paper, we investigate the stellar parameters of $\epsilon$~CMa and the attenuation of its 
ionizing spectrum.  Using non-LTE, line-blanketed model atmospheres for the stellar EUV continuum
and flux decrement at the Lyman limit (LL), we determine the attenuation of the EUV in both the 
stellar atmosphere and ISM.  In Section 2, we derive a new set of stellar parameters (radius, mass, 
effective temperature, luminosity) consistent with measurements of the parallax distance, stellar 
angular diameter, and total flux.   In Section 3, we analyze the absorption in the stellar atmosphere 
and intervening ISM.  We find an interstellar \HI\ column density 
$N_{\rm HI} = (6\pm1)\times10^{17}~{\rm cm}^{-2}$, corresponding to a LL optical depth 
$\tau_{\rm LL} = 3.8\pm0.1$ and a flux decrement of a factor of $\Delta_{\rm ISM} = 45\pm5$.  
Combined with a decrement $\Delta_{\rm star} = 110\pm10$ in the stellar atmosphere, the total 
decrement $\Delta_{\rm LL} = 5000$ is consistent with the EUV observations.  In Section 4, we 
discuss the implications of our results for the ionization structure of the local interstellar clouds.

\section{Revised Stellar Parameters}

\subsection{Implications of the Closer Distance} 

A key measurement for $\epsilon$~CMa is its angular diameter $\theta_d = 0.80\pm0.05$~mas 
(R.\ Hanbury Brown \etal\ 1974) using the stellar interferometer at the Narrabri Observatory.  From 
this and the parallax distance $d = 124\pm2$~pc, we derive a stellar radius of
\begin{equation}
     R = \left( \frac {\theta_d \, d} {2} \right) = 7.42\times10^{11}~{\rm cm}  
           ~~ (10.7\pm0.7~R_{\odot})   \; , 
\end{equation}  
where we combined the relative errors on $\theta_d$ (6.3\%) and $d$ (1.7\%) in quadrature.  
This  ``interferometric radius" is considerably smaller than previous values in the literature, and
the revision in stellar distance from 188~pc to 124~pc will alter many stellar properties.  

The first change is the reduction in stellar radius to $10.7~R_{\odot}$, previously listed as 
$16.6~R_{\odot}$ (T.\ Snow \& D.\ Morton 1976) and $16.2\pm1.2~R_{\odot}$ (J.\ Cassinelli \etal\ 1995).   
L.\ Fossati \etal\ (2015) adopted $d = 124\pm2$~pc and evaluated the radius and mass from two sets 
of evolutionary models:  $R = 12.0^{+1.7}_{-1.5}~R_{\odot}$ and $M = 13.1^{+1.0}_{-0.9}~M_{\odot}$ 
(tracks from C.\ Georgy \etal\ 2013) and $R = 10.1^{+0.7}_{-0.5}~R_{\odot}$ and 
$M = 12.0^{+0.4}_{-0.4}~M_{\odot}$ (tracks from L.\ Brott \etal\ 2011).  The new parallax distance implies 
a fainter absolute magnitude and  suggests that its luminosity class might be II/III (giant) rather than 
II (bright giant) derived from its spectral morphology (L.\ Fossati \etal\ 2015; I.\ Negueruelo \etal\ 2024).
This would be consistent with its smaller interferometric radius.

A second change concerns the elevated EUV fluxes from $\epsilon$~CMa. J.\ Cassinelli \etal\ (1995)
noted that these fluxes are an order of magnitude larger than predicted by model stellar atmospheres, 
both LTE (R.\ Kurucz 1979) and non-LTE (I.\ Hubeny \etal\ 1994).  
Using high-resolution spectra, L.\ Fosatti \etal\ (2015) changed the stellar classification from B2~II 
to B1.5~II, raised the effective temperature from $T_{\rm eff} = 20,990\pm760$~K (A.\ Code \etal\ 1976) 
to $22,500\pm300$~K, and increased the surface gravity from $\log g = 3.20$ to $3.40\pm0.08$.   
A recent study of spectral classifications of B-type stars (I.\ Negueruela \etal\ 2024) also listed 
$\epsilon$~CMa as B1.5~II, based on morphological comparison of its optical spectrum with other
B-star standards. However, as discussed below (Section 2.3), the higher $T_{\rm eff}$ is inconsistent 
with the bolometric relation between stellar integrated flux and surface area, 
$T_{\rm eff} = (L/4 \pi R^2 \sigma_{\rm SB})^{1/4}$.  A value $T_{\rm eff} = 21,000$~K is more 
consistent with the bolometric luminosity and radius.  However, because of the back-warming effects 
of a stellar wind, a model-atmosphere temperature of $22,500$~K may be needed to interpret the
elevated EUV continuum fluxes.  In Section~2.3 we examine the stellar parameters 
($M$, $R$, $g$, $T_{\rm eff}$) required for consistency with the integrated flux $f$, luminosity $L$, 
and bolometric magnitude $M_{\rm bol}$.


\begin{figure}[ht]
\includegraphics[angle=0,scale=0.90] {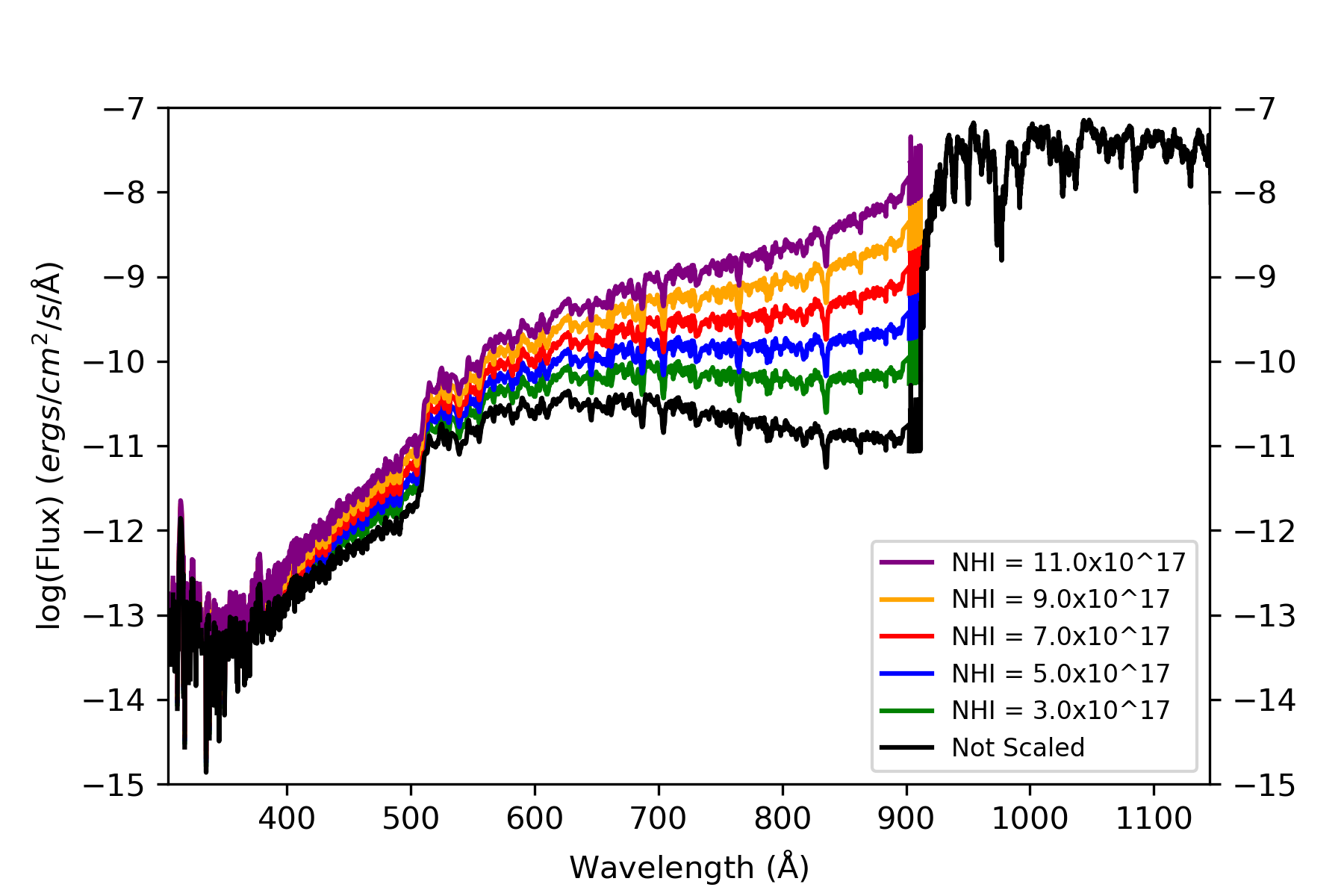}
\caption{The lowest curve (black) shows the flux-calibrated spectrum of $\epsilon$~CMa 
(N.\ Erickson \etal\ 2021), a combination of data from the DEUCE rocket flight and previous 
EUVE and IUE data to which DEUCE was scaled.  The five higher curves show restored 
continua for column densities $N_{\rm HI} =$ (3--11)$\times10^{17}~{\rm cm}^{-2}$ and
$N_{\rm HeI} = 0.06\,{\rm N}_{\rm HI}$ based on observed elevated He$^+$ ionization
fractions.  These curves multiply the observed fluxes by $\exp(\tau_{\lambda})$ for
optical depths $\tau_{\lambda}$ in the ionizing continua of \HI\ ($\lambda \leq 912$~\AA) 
and \HeI\ ($\lambda \leq 504$~\AA). The factor of $5000\pm500$ flux decrement at 
912~\AA\ (black curve) is a combination of a ($110\pm10$) decrement in the stellar 
atmosphere and a ($45\pm5$) decrement from interstellar absorption. We then infer 
an ISM optical depth $\tau_{\rm LL} = 3.8\pm0.1$ at the 912~\AA\ edge and 
$N_{\rm HI} = (6\pm1) \times10^{17}$~cm$^{-2}$.}
\end{figure}
 

\subsection{Absolute Magnitude and Classification of $\epsilon$~CMa.}  

The absolute magnitude and bolometric magnitude of $\epsilon$~CMa need revision, consistent 
with its closer distance $d = 124\pm2$~pc.  The observed visual magnitude $m_V = 1.50$ 
corresponds to absolute magnitude $M_V = -3.97\pm0.04$. The stellar classification of B1.5~II 
(L.\ Fossati \etal\ 2015) and $M_V = -5.1$ from J.\ Lesh (1968) interpolated to B1.5~II results in a large
distance discrepancy, with ($m_V$--$M_V) = 6.6$ and $d = 209$~pc.  In a compilation of absolute 
magnitudes for OB stars (Table 11 in D.\ Bowen \etal\ 2008) the recommended values of
$M_V = -4.7$ (B2~II) and $M_V = -4.9$ (B1.5~II) differ slightly from those in J.\ Lesh (1968).
However, these also imply unacceptably large distances:  ($m_V$--$M_V) = 6.20$ ($d = 173$~pc) 
for B2~II  and ($m_V$--$M_V) = 6.40$ ($d = 190$~pc) for B1.5~II. 

Absolute magnitudes change rapidly with luminosity class at spectral types B1--B2. Classifications 
of B2~II ($M_V = -4.9$) and B2~III ($M_V = -3.3$) straddle the observational value of $\epsilon$~CMa 
($M_V \approx -4.0$).  We therefore suggest that $\epsilon$~CMa may be luminosity class B2~II/III 
rather than B2~II, providing a consistent bolometric magnitude with distance modulus 
($m_V$--$M_V) = 5.467$ ($d = 124$~pc) and $M_V = -3.97$.   For reference, the solar absolute 
bolometric magnitude $M_{{\rm bol},\odot} = 4.74$ corresponds to luminosity 
$L_{\odot} = 3.828\times10^{33}~{\rm erg~s}^{-1}$.  Figure~5 in M.\ Pedersen \etal\ (2020) 
shows a bolometric correction ($M_{\rm bol}$--$M_V) = -2.0$ for $\log g = 3.5$ and 
$T_{\rm eff} = 21,000$~K, leading to a bolometric absolute magnitude $M_{\rm bol} = -5.97$  
and $L =  19,200~L_{\odot}$.  In contrast, the bolometric correction at $T_{\rm eff} = 22,500$~K
is $-2.2$, yielding $M_{\rm bol} = -6.17$  and $L =  23,000~L_{\odot}$.  The 16\% larger luminosity
arises from a mismatch between the bolometric temperature and the elevated temperature 
(22,500~K) invoked to produce the EUV continuum.  

\subsection{Stellar Luminosity and Effective Temperature} 

The discrepancy between spectrophotometric and parallax distances requires 
revising $M_V$ and $M_{\rm bol}$ for $\epsilon$~CMa.  Consistent values of $R$, $T_{\rm eff}$, 
$M_{\rm bol}$, and $L$ must satisfy the relations $g =  (G M/R^2)$ and 
$L = 4 \pi R^2 \sigma_{\rm SB} T_{\rm eff}^4$, where $\sigma_{\rm SB} = 5.6704 \times 10^{-5}$
erg~cm$^{-2}$~s$^{-1}$~K$^{-4}$ is the Stefan-Boltzmann constant and $R = (\theta_d \, d/2)$.  
Here, $\theta_d = 3.88 \times10^{-9}$~rad ($0.80\pm0.05$ mas) is the measured stellar angular 
diameter (R.\ Hanbury Brown \etal\ 1974).  The integrated stellar flux 
is $f = (41.5 \pm 3.3) \times10^{-6}~{\rm erg~cm}^{-2}~{\rm s}^{-1}$ from A.\ Code \etal\ (1976), 
who combined ground-based visual and near-infrared photometry with ultraviolet observations 
(1100--3500~\AA) taken by the Orbiting Astronomical Observatory (OAO-2).  This flux included 
an extrapolated value for shorter wavelengths 
$f(\lambda \leq 1100~{\rm \AA}) = 6.5\times10^{-6}~{\rm erg~cm}^{-2}~{\rm s}^{-1}$,
in agreement with the flux-calibrated DEUCE rocket observations from 700--1150~\AA\ 
(N.\ Erickson \etal\ 2021).   Even after correcting the observed EUV flux for interstellar
absorption, the Lyman continuum represents less than 1\% of the bolometric flux.  

The constraints from stellar angular diameter and total flux give a useful relation for the 
bolometric effective temperature $T_{\rm eff} = (4f / \sigma_{\rm SB} \, \theta_d^2)^{1/4}$ or  
\begin{equation} 
    T_{\rm eff} = (21,000\pm780~{\rm K}) \left( \frac {f} {41.5\times10^{-6}} \right)^{1/4} 
             \left( \frac {\theta_d} {0.80~{\rm mas}} \right)^{-1/2}       \;  . 
\end{equation}
From the stellar radius and bolometric $T_{\rm eff}$, we find the total luminosity,
\begin{equation}
   L = (19,900~L_{\odot}) \left( \frac {T_{\rm eff}}{21,000~{\rm K}} \right)^4 
          \left( \frac {R}{10.7~R_{\odot}} \right)^2  \; ,
\end{equation}
which agrees by construction with the relation $L = 4 \pi d^2 f$.  With the relative errors
on parallax distance (1.7\%) and integrated flux (8\%), the luminosity is known to 8.2\%.  
Given the accuracy of the integrated flux and distance, we will maintain the bolometric value
$T_{\rm eff} = 21,000$~K.  However, as discussed in Section 2.4, the elevated model 
temperature of 22,500~K from stellar-wind backwarming may provide a better fit to the the 
EUV continuum and stellar flux decrement at the Lyman limit.  Table 1 summarizes the 
stellar parameters used in previous papers, together with our revised values.   

\subsection{Stellar Atmospheres and EUV Fluxes}

Two problems remain in understanding the elevated EUV fluxes from $\epsilon$~CMa and in
reconciling the two values of $T_{\rm eff}$ (21,000~K and 22,500~K) frequently quoted in the literature.  
J.\ Cassinelli \etal\ (1995) noted that  ``the emergent flux from $\epsilon$~CMa in the hydrogen Lyman 
continuum exceeds the value expected from model atmospheres by a factor of about 30, even when 
the effects of line blanketing are accounted for."  Because this discrepancy was found in both LTE 
and non-LTE model atmospheres with $T_{\rm eff} = 21,000$~K, they suggested that back-warming 
by the shocked stellar wind could boost the temperature in the upper atmosphere where the Lyman 
continuum is formed.  A similar non-LTE wind effect was proposed by F.\ Najarro \etal\ (1996), involving 
doppler shifts and velocity-induced changes in density that affect the escape of \HI\ and \HeI\
resonance lines and their ground-state populations.  Both mechanisms are sensitive to
mass-loss rates in the range 
$\dot{M} \approx$ (1--10)$\times10^{-9}~M_{\odot}~{\rm yr}^{-1}$.  

J.\ Cassinelli \etal\ (1995) compared the observed fluxes of $\epsilon$~CMa to model atmospheres 
with $T_{\rm eff} = 21,000$~K and $\log g = 3.20$ ($g = 1585~{\rm cm~s}^{-2}$) and adopted 
$M = 15.2~M_{\odot}$ and $R = 16.2~R_{\odot}$.   In their spectroscopic analysis of 
$\epsilon$~CMa, L.\ Fosatti \etal\ (2015) derived $\log g = 3.40\pm0.08$ and spectral type B1.5~II. 
In this paper, for consistency with the parameter changes (radius, distance, interferometric diameter),
we increased $\log g$ to $3.50\pm0.05$ ($g = GM/R^2 = 3162~{\rm cm~s}^{-2}$). The rotational 
velocity has been measured at $V_{\rm rot} \sin i = 21.2\pm2.2$~\kms\ (Fosatti \etal\ 2015). 
With $R = 10.7\pm0.7~R_{\odot}$, we can safely neglect the centrifugal term,
$V_{\rm rot}^2/R \approx (6~{\rm cm~s}^{-2})(\sin i)^{-2}$ and infer a gravitational mass 
($M = gR^2/G$) of  
\begin{equation}
   M = (13.1\pm2.3~M_{\odot}) \left( \frac {g}{3162~{\rm cm~s}^{-2}} \right) 
        \left( \frac {R} {10.7~R_{\odot}} \right)^2   \; .
\end{equation}
The quoted error on $M$ includes uncertainties in surface gravity and radius, added in quadrature 
with $(\sigma_M/M)^2 = (\sigma_g/g)^2 + (2 \sigma_R/R)^2$.   

The position of $\epsilon$~CMa on the Hertzsprung-Russell (H-R) diagram is shown in Figure 2, 
with our revised parameters, $\log (L/L_{\odot}) = 4.30$ and $T_{\rm eff} = 21,000$~K, and evolutionary 
tracks from Brott \etal\ (2011).   Its position falls close to the $12~M_{\odot}$ track, not far from
our mass estimate in eq.\ (4) with $R = 10.7~R_{\odot}$ and $\log g = 3.5$.  A lower value 
($\log g = 3.40\pm0.08$) would reduce the gravitational mass to $10.4\pm2.5~M_{\odot}$.
This H-R position is consistent with luminosity class~II/III (giant) rather than class~II (bright giant). 
It also places $\epsilon$~CMa out of the $\beta$~Cephei instability strip, consistent 
with the lack of observed pulsations. This conclusion was confirmed by comparison with the locus of 
radial and non-radial instability modes shown in Figures 2 and 3  of L.\ Deng \& D.~R.\ Xiong (2001).


\begin{figure}[ht]
\includegraphics[angle=0,scale=0.90] {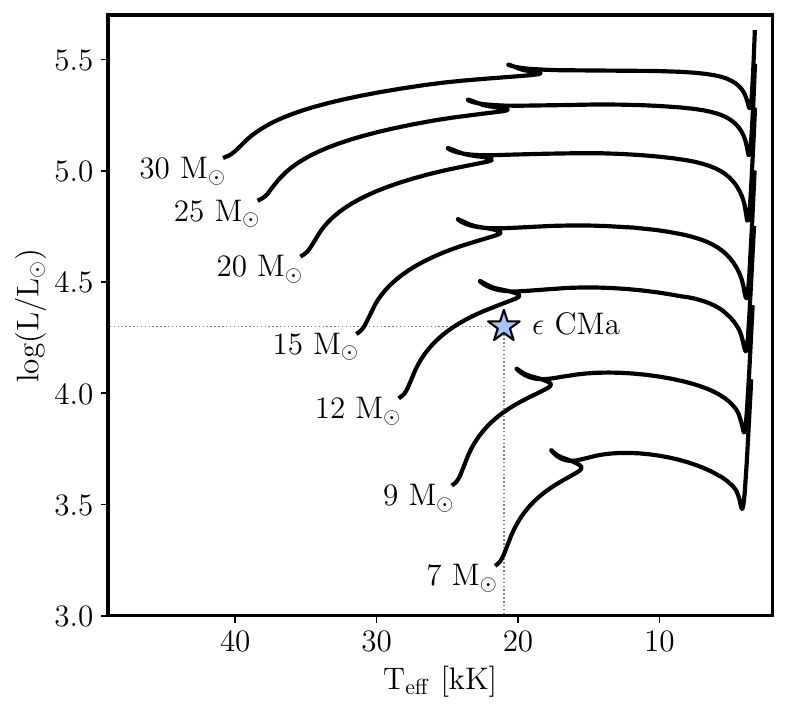}
\caption{The location of $\epsilon$~CMa on the Hertzsprung-Russell diagram is shown 
for our revised parameters, $\log (L/L_{\odot}) = 4.30$ and $T_{\rm eff} = 21,000$~K, 
based on new radius $R = 10.7~R_{\odot}$ (for $d = 124$~pc).  The evolutionary tracks 
are from Brott \etal\ (2011) with Milky Way metallicities and initial masses labeled from 
7--30 $M_{\odot}$. 
 }
\end{figure}

\section{Interstellar Absorption and \HI\ Column Density } 

\subsection{Previous Column Density Estimates}

Figure 1 shows the flux-calibrated FUV and EUV spectra of $\epsilon$~CMa taken 
by EUVE (J.\ Dupuis \etal\ 1995) and the Colorado--DEUCE rocket (N.\ Erickson \etal\ 2021).
The continuum flux drops by a factor $\Delta_{\rm LL} = 5000\pm500$ at the 912~\AA\  
Lyman limit,  with a further decrease at the \HeI\ photoionization edge (504~\AA).  
The total optical depth at 912~\AA\ is $\tau_{\rm LL} = \ln \Delta_{\rm LL}$.  
Some attenuation is intrinsic to the stellar atmosphere, owing to \HI\ opacity that 
varies with SpT and $T_{\rm eff}$.  Additional photoelectric absorption occurs in the ISM. 
The fact that $\epsilon$~CMa is a strong source of local EUV radiation is largely a result 
of its location in a low-density cavity or local interstellar tunnel.

Past estimates for the amount of interstellar hydrogen toward $\epsilon$~CMa span
a wide range (Table 2).  Using {\it Copernicus} ultraviolet spectral fits to the wings of the 
interstellar \Lya\ absorption line, R.\ Bohlin (1975) set an upper limit of 
$N_{\rm HI} < 5\times10^{18}~{\rm cm}^{-2}$, corresponding to optical depth 
$\tau_{\rm LL} < 31.5$.  The actual ISM column density is much less, since detectable
EUV flux makes it through the local clouds to Earth.   Because \Lya\ absorption measurements 
are uncertain at low \HI\ column densities, several groups have estimated $N_{\rm HI}$ by scaling 
from column densities of  heavy elements believed to co-exist with \HI.  Using ultraviolet 
spectra taken with the G160M grating (resolving power $R \approx 20,000$) of the Goddard
High Resolution Spectrograph (GHRS) aboard the Hubble Space Telescope (HST),  
C.\ Gry \etal\ (1995) estimated that $N_{\rm HI} < 5\times10^{17}~{\rm cm}^{-2}$ scaled from 
\NI\ absorption lines.  This estimate required assumptions about the gas-phase metallicities 
and possible corrections for depletion onto dust grains.  They used a mean interstellar nitrogen 
abundance of $10^{-4.17}$ relative to hydrogen found by R.\ Ferlet (1981) in unreddened sight 
lines observed by {\it Copernicus}.  They also noted difficulties produced by blending of the 
interstellar \NI\ (1199--1200~\AA) triplet absorption lines with narrow stellar features.  

An improved analysis by C.\ Gry \& E.\ Jenkins (2001) used HST observations of ultraviolet 
absorption lines of \NI, \OI, and \SII\  taken with the echelle grating of GHRS ($R\sim100,000$).  
The increased spectral resolution and high signal-to-noise ratio of these data allowed them to 
identify three principal velocity components at heliocentric velocities 17, 10, and $-10$ \kms, 
previously referred to as Components 1, 2, and 3.  Component~1 is known as the 
Local Interstellar Cloud (LIC) and Component~2 is called the Blue Cloud, also seen in 
absorption toward Sirius ($d = 2.64$~pc).  Scaling from \NI, they estimated a range of
\HI\ column densities $N_{\rm HI} =$ (3.1--3.7)$\times10^{17}~{\rm cm}^{-2}$.   They 
quoted two estimates scaled from \OI, with
$N_{\rm HI} =$ (7--11)$\times10^{17}~{\rm cm}^{-2}$ for O/H = $3.16\times 10^{-4}$ and
$N_{\rm HI} =$ (4.8--7.2)$\times10^{17}~{\rm cm}^{-2}$ for O/H = $4.68\times 10^{-4}$.
The difference in these estimates (see Table 2) reflects their assumptions about (N/H and O/H) 
abundances and possible depletion into grains.  The offset between N and O scalings could
indicate sub-solar interstellar nitrogen abundances.  A \NI\ deficiency in the local ISM was 
noted by E.\ Jenkins \etal\ (2000) from far-UV spectra of white dwarf stars observed with FUSE.  
However, in a survey of B-type stars, M.-F.\ Nieva \& N.\ Przybilla (2012) noted that 
$\epsilon$~CMa had a slightly elevated stellar N/H abundance (0.2--0.3 dex) compared to 
unmixed B~stars in the solar neighborhood.  

\subsection{Photoelectric Absorption} 

In this section, we derive $N_{\rm HI}$ by comparing the observed EUV continuum fluxes
to estimates of the stellar continuum (non-LTE model atmospheres) and attenuation by the ISM.  
We restore the observed continuum to its shape at the stellar surface by multiplying the observed 
flux by $\exp(\tau_{\lambda})$, using optical depths $\tau(\lambda)$ of photoelectric absorption 
in the ionizing continua of \HI\  ($\lambda \leq 912$~\AA) and \HeI\ ($\lambda \leq 504$~\AA),
\begin{eqnarray}
 \tau_{\rm HI} (\lambda) &\approx& (0.630) \left( \frac {N_{\rm HI} } {10^{17}~{\rm cm}^{-2} } \right) 
           \left( \frac {\lambda} {912~{\rm \AA} } \right)^3   \; , \\
 \tau_{\rm HeI} (\lambda) &\approx& (0.737) \left( \frac {N_{\rm HeI} } {10^{17}~{\rm cm}^{-2} } \right) 
           \left(  \frac {\lambda} {504~{\rm \AA} } \right)^{1.63}   \; .   
\end{eqnarray}
These approximations are based on power-law fits to the photoionization cross sections at 
wavelengths below threshold,
$\sigma_{\rm HI}(\lambda) \approx (6.30\times10^{-18}~{\rm cm}^{2})(\lambda/912~{\rm \AA})^3$
(D.\ Osterbrock \& G.\ Ferland 2006) and 
$\sigma_{\rm HeI}(\lambda) \approx (7.37\times10^{-18}~{\rm cm}^{2})(\lambda/504~{\rm \AA})^{1.63}$
(D.\ Samson \etal\ 1994).  In our actual calculations, we used the exact (non-relativistic) \HI\ 
cross section (Bethe \& Salpeter 1957), which has frequency dependence,
\begin{equation} 
   \sigma_{\nu} = \sigma_0 \left( \frac {\nu}{\nu_0} \right)^{-4} 
             \frac {\exp[ 4 - (4 \arctan \epsilon ) / \epsilon ]}  { [1 - \exp(-2 \pi / \epsilon) ] }  \; .
\end{equation}
Here, the dimensionless parameter $\epsilon \equiv [(\nu/\nu_0) -1]^{1/2}$ with frequency $\nu_0$ 
defined at the ionization energy $h \nu_0 = 13.598$~eV and 
$\sigma_0 =  6.304\times10^{-18}~{\rm cm}^{2}$.  The two formulae agree at threshold $\nu = \nu_0$, 
but the approximate formula deviates increasingly at shorter wavelengths.  The exact cross section is 
higher by 8.2\% (700~\AA), 12.3\% (600~\AA), and 16.4\% (500~\AA).   In our models, we found that 
the exact cross section increased the photoionization rates by 10--13\%  in models with
$N_{\rm HI} =$ (3--8)$\times 10^{17}~{\rm cm}^{-2}$.

The flux-restoration method for inferring ISM opacities from the observed EUV flux has inherent 
uncertainties.  It requires accurate model-atmosphere estimates of the flux decrement factor
$\Delta_{\rm star}$ at the Lyman limit  and the shape of the EUV continuum below the ionization 
edge.   Especially important are non-LTE treatments of the \HI\ and \HeI\ abundances in their 
(1s and 1s$^2$) ground states  and possible back-warming effects of stellar winds.  
J.\ Cassinelli \etal\ (1995) noted that EUVE fluxes in both \HI\ and \HeI\ continua were significantly
greater than predicted by both LTE and non-LTE model atmospheres.  They quoted a wide range of 
$N_{\rm HI} =$ (0.7--1.2)$\times 10^{18}~{\rm cm}^{-2}$.

To improve on this technique, we followed a similar procedure, calculating the interstellar opacities 
of both \HI\ and \HeI.  In the flux-calibrated data (Figure~1) the far-UV stellar continuum rises slowly 
from 1000~\AA\ down to 912~\AA, owing to absorption in higher Lyman-series lines converging on 
the LL at 911.75~\AA.  We estimate the extrapolated continuum level at $912^+$~\AA, longward 
of the edge, as $F_{\lambda} = 5\times10^{-8}~{\rm erg~cm}^{-2}~{\rm s}^{-1}~{\rm \AA}^{-1}$, 
dropping by a factor $\Delta_{\rm LL} \approx 5000\pm500$ to 
$1.0\times10^{-11}~{\rm erg~cm}^{-2}~{\rm s}^{-1}~{\rm \AA}^{-1}$ just below the Lyman edge.   

We compared these restored continua to the EUV continuum shape and the Lyman flux decrement
in model atmospheres, using the code 
\texttt{WM-basic} developed by A.\ Pauldrach \etal\ (2001)\footnote{This code can be found at 
http://www.usm.uni-muenchen.de/people/adi/Programs/Programs.html}.
We chose this code because of its hydrodynamic solution of expanding atmospheres with 
line blanketing and non-LTE radiative transfer, including its treatment of the continuum and
wind-blanketing from  EUV lines.   This code was the standard for atmosphere modeling in the 
population synthesis code \texttt{Starburst99} (C.\ Leitherer \etal\ 1999, 2014) and used by 
M.\ Topping \& J.~M.\ Shull (2015) to evaluate the production rate of  Lyman continuum 
radiation in OB-stars.   Extensive discussion of hot-star atmosphere codes appears in papers 
by D.~J.\ Hillier \& D.\ Miller (1998), F.\ Martins \etal\ (2005), and C.\ Leitherer \etal\ (2014).  

\subsection{Combined Stellar and Interstellar Absorption}

The restored EUV continua shown in Figure 1 covered the range of interstellar \HI\ column 
densities, (3--11)$\times10^{17}$~cm$^{-2}$, quoted in the literature.  After correcting for 
photoelectric absorption of the EUVE and DEUCE fluxes, we found that values 
$N_{\rm HI} \geq 9 \times10^{17}~{\rm cm}^{-2}$ are inconsistent with the observed flux 
decrements at the Lyman edge, based on model atmospheres and using the exact \HI\ 
cross section in equation (7).  We narrowed our study to interstellar \HI\ column densities 
in the range (5--8)$\times10^{17}~{\rm cm}^{-2}$.

The total flux decrement factor $\Delta_{\rm LL}$ at the Lyman limit is the combination of 
attenuation in the stellar atmosphere ($\Delta_{\rm star}$) and the interstellar medium 
($\Delta_{\rm ISM}$).   We constrain the product of stellar and interstellar attenuation
factors by finding the optical depth ($\tau_{\rm ISM} = \ln \Delta_{\rm ISM}$) at the 912~\AA\ 
edge consistent with the range of LL decrements found in a set of model atmospheres run 
for B1.5--B2 giants ($T_{\rm eff} =$ 21,000--22,500~K and $\log g =$ 3.4--3.5).  Because  
LL optical depth depends logarithmically on the \HI\ column density, values of $\tau_{\rm LL}$
provide better constraints on $N_{\rm HI}$ than broad-band fits between 504-912~\AA.  As 
described below, we adopted a combination of $\Delta_{\rm star} = 110\pm10$ and 
$\Delta_{\rm ISM} = 45\pm5$ to produce a total decrement of 
$\Delta_{\rm LL} = \Delta_{\rm star}  \times \Delta_{\rm ISM} = 5000 \pm 500$.

Figure~3 shows the FUV/EUV fluxes and Lyman flux decrements for three non-LTE, 
line-blanketed model atmospheres.  Two models adopt $T_{\rm eff} = 22,500$~K, the elevated 
temperature proposed  by L.\ Fossati \etal\ (2015), with $\log g = 3.40$ and 3.50.  A third model 
assumes $T_{\rm eff} = 21,000$~K and $\log g = 3.50$, consistent with the bolometric relations 
between integrated flux and angular diameter (see eq.\ 2).  The hotter models with stellar winds
include back-warming of the star's outer atmosphere, which raises the temperatures where the 
emergent ionizing continua of \HI\ and \HeI\ are produced.  The two 22,500~K (\texttt{WM-basic}) 
models exhibit LL decrements $\Delta_{\rm star}$ of 102 and 119.   We also analyzed decrements 
in two CMFGEN model atmospheres (D.~J.\ Hillier \& D.\ Miller 1998) with $T_{\rm eff} = 22,500$~K 
and $\log g = 3.50$, provided by D.~J. Hillier (private communication).  These models
include stellar winds with mass-loss rates of 1--4 $\times10^{-9}~M_{\odot}~{\rm yr}^{-1}$ and
produce LL decrements $\Delta_{\rm star} \approx 120-140$.


\begin{figure*}[ht]
\includegraphics[angle=0,scale=0.70] {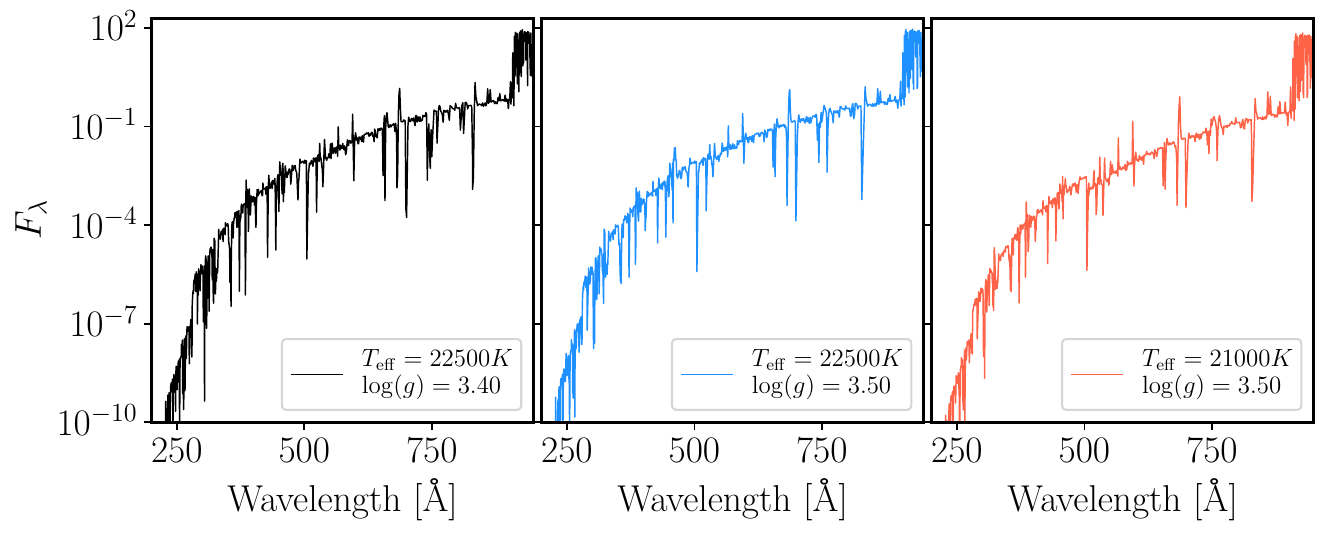}
\includegraphics[angle=0,scale=0.71] {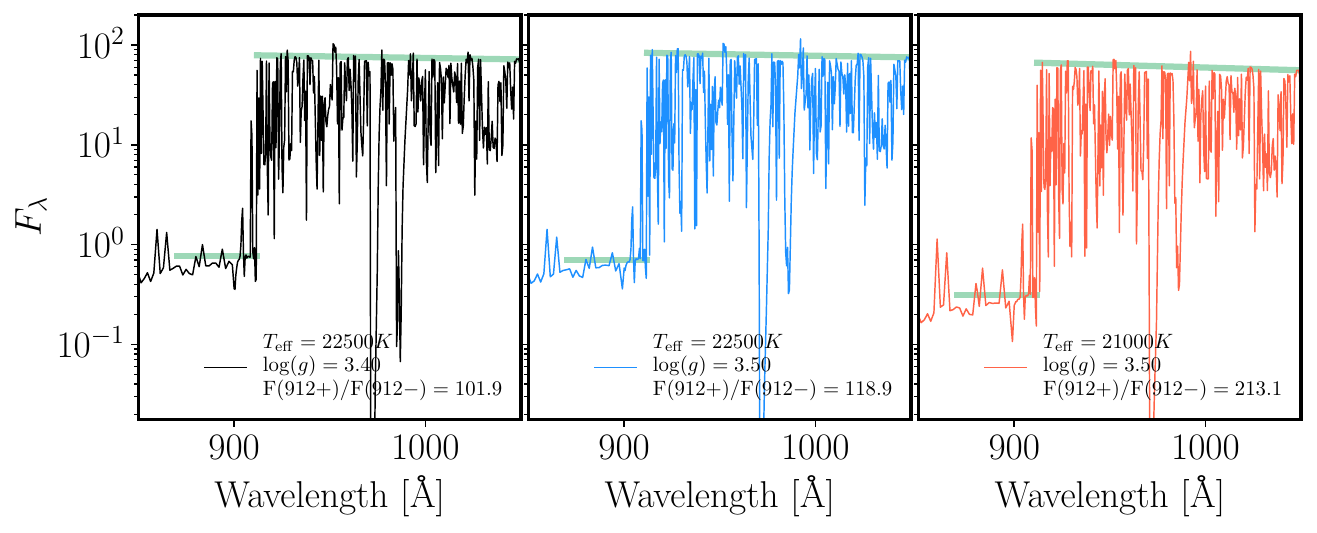}
\caption{(Top panels).  Far-UV and EUV spectra from three model atmospheres for 
$\epsilon$~CMa, computed with the non-LTE line-blanketed code \texttt{WM-basic} 
and plotting the flux distribution $\log F_{\lambda}$.  Two models use the elevated 
effective temperature $T_{\rm eff} = 22,500$~K  and surface gravities $\log g = 3.40$ 
and 3.50.  A third model uses $T_{\rm eff} = 21,000$~K based on integrated-flux 
bolometry (A.\ Code \etal\ 1976).  The absence of a \HeI\ edge (504~\AA) may result 
from back-warming of the upper atmosphere by a line-blanketed stellar wind.
(Bottom panels.)  Zoom-in plots of flux decrements $F(912^+)/F(912^-)$ at the  
Lyman edge.  These factors are $\Delta_{\rm LL} = 102$ and 119 for the two 
22,500~K models and $\Delta_{\rm LL} = 213$ for the 21,000~K model.   }
\end{figure*}

We adopt a stellar decrement  $\Delta_{\rm star} = 110\pm10$ based on \texttt{WM-basic} models 
(see Figure~3).  To produce the observed total Lyman decrement $\Delta_{\rm LL} = 5000\pm500$,
the ISM generate an additional decrement $\Delta_{\rm ISM} = 45\pm5$, 
corresponding to optical depth $\tau_{\rm LL} = 3.8\pm0.1$ and \HI\ column density 
N$_{\rm HI} = (6 \pm 1) \times 10^{17}~{\rm cm}^{-2}$.   We estimate 10\% uncertainty in defining 
the continuum longward of 912~\AA\ and 15\% uncertainty in the LL decrement 
in the model atmospheres.   Because of the weak scaling of optical depth with flux decrement
($\tau_{\rm LL} = \ln \Delta_{\rm LL}$) the uncertainties on $\tau_{\rm ISM}$ and column 
density N$_{\rm HI}$ are small. Figure 4 shows the observed and restored EUV continua for 
our constrained \HI\ column density, overlaid with the stellar atmosphere model with 
decrement $\Delta_{\rm LL} = 102$.  The combined stellar and interstellar decrements are in good 
agreement with the flux-restored EUV spectrum from 912~\AA\ down to 750~\AA.  The model 
atmosphere falls below the restored continuum from 750~\AA\ down to 504~\AA, but it 
fits the \HeI\ continuum quite well at $\lambda < 504$~\AA.


\begin{figure}[ht]
\includegraphics[angle=0,scale=1.1] {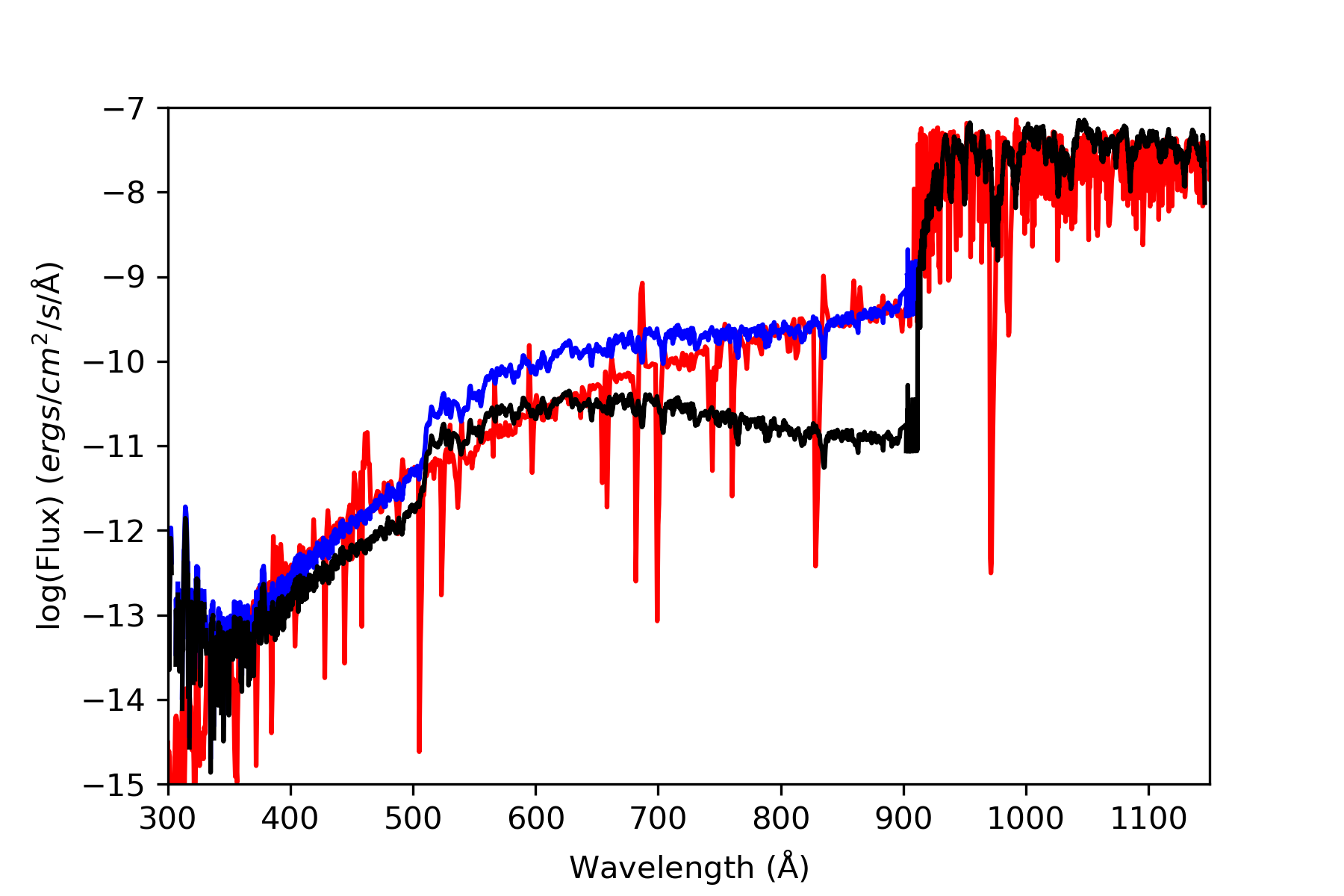}
\caption{We illustrate the combined attenuation of the EUV continuum of $\epsilon$~CMa
by both the stellar atmosphere and ISM.  The red curve shows the \texttt{WM-basic} model 
atmosphere spectrum ($T_{\rm eff} = 22,500$~K and $\log g = 3.50$) overlaid on the 
observed EUV/FUV spectrum (black curve).  The flux-restored spectrum (blue) assumes 
an intervening \HI\ column density $N_{\rm HI} = 6\times10^{17}~{\rm cm}^{-2}$, with
optical depth $\tau_{\rm LL} = 3.8\pm0.1$ at the Lyman edge.  The observed spectrum has 
a total LL flux decrement $\Delta_{\rm LL} = F(912^+)/F(912^-) = 5000\pm500$.  
Our model has a stellar decrement $\Delta_{\rm star} = 110\pm10$ and an 
ISM decrement $\Delta_{\rm ISM} = 45\pm5$.
}
\end{figure}
 

 Table 3 summarizes the results of our models of the restored ionizing continua for a range of column
 densities.  These parameters include the integrated photon flux $\Phi_{\rm LyC}$, photoionization 
 rate $\Gamma_{\rm HI}$, Lyman decrement $\Delta_{\rm ISM}$, and the hydrogen 
 ionization fraction at the outer surface of the local cloud.  In photoionization equilibrium, 
 these fractions, $x = (1/2)[ -a + (a^2 + 4a)^{1/2}]$, are the solution of $x^2/(1-x) = a$, where 
 $a = \Gamma_{\rm H} / (1.1 n_{\rm H} \alpha_{\rm H})$.
We included a factor 1.1 for electrons contributed by He$^+$, although helium may be somewhat 
more ionized than hydrogen owing to contributions from hot white dwarfs and EUV emission lines
produced in the hot local bubble. The local cloud is assumed to have constant hydrogen density 
$n_{\rm H} \equiv n_{\rm HI} + n_{\rm HII} \approx 0.2~{\rm cm}^{-3}$ with a hydrogen case-B 
radiative recombination coefficient 
$\alpha_{\rm H} = 3.39\times10^{-13}~{\rm cm}^3~{\rm s}^{-1}$ at $T = 7000$~K. 
For our best fit, $N_{\rm HI} = (6\pm1)\times10^{17}$ cm$^{-2}$, the EUV radiation 
field outside the local clouds is 10--20 times higher than viewed from Earth, 
with $\Phi_{\rm LyC} \approx 3000\pm1000$ photons~cm$^{-2}~{\rm s}^{-1}$ and  
$\Gamma_{\rm H} =$ (1--2)$\times10^{-14}~{\rm s}^{-1}$.  At the outer surface of the local 
clouds exposed to the higher LyC flux, hydrogen is partially ionized, with ionization fractions 
$x_{\rm s} \approx$~30--40\% for $n_{\rm H} = 0.2$~cm$^{-3}$.  Somewhat higher fractions
are found if we reduce the density to $n_{\rm H} = 0.1~{\rm cm}^{-3}$. 

\section{Summary of Results and Future Studies}  

We have derived a new set of stellar parameters for $\epsilon$~CMa (mass, radius, effective 
temperature, luminosity) consistent with its shorter parallax distance (124~pc vs.\ 188~pc), 
interferometric angular diameter ($\theta_d = 0.80\pm0.05$~mas), and integrated bolometric 
flux, $f = (41.5 \pm 3.3) \times10^{-6}~{\rm erg~cm}^{-2}~{\rm s}^{-1}$.  From these, we derive
$T_{\rm eff} = 21,000\pm780$~K and $L \approx 20,000~L_{\odot}$.   With updated absolute 
magnitudes ($M_V = -3.97\pm0.04$ and $M_{\rm bol} = -5.97$) $\epsilon$~CMa is 
sub-luminous by 0.9--1.1 mag for its morphological classification as a bright giant,
either B2~II (Lesh 1972) or B1.5~II (Fossati \etal\ 2015), and it has inconsistent values of 
$R$, $T_{\rm eff}$, $d$, and radiative flux.  On the H-R diagram, the new parameters 
shift $\epsilon$~CMa out of the $\beta$-Cephei pulsational instability strip, consistent 
with its observed lack of pulsations and the boundaries of theoretical instability on evolutionary 
tracks (L.\ Deng \& D.~R.\ Xiong 2001).

\noindent
The following points summarize our primary results:
\begin{enumerate}

\item The combination of parallax distance and angular diameter yield a stellar radius 
$R = 10.7\pm0.7~R_{\odot}$ and luminosity $L \approx 20,000~L_{\odot}$, both smaller
 than previous estimates.  Bolometric relations between  flux and radius 
suggest an effective temperature $T_{\rm eff} \approx 21,000\pm780$~K and absolute
magnitude $M_V = -3.97\pm0.04$, appropriate for a B2~II/III star.  A previous spectroscopic
analysis (Fosatti \etal\ 2015) gives $T_{\rm eff} = 22,500$~K at SpT = B1.5~II.  

\item From models of the stellar and interstellar attenuation of the ionizing flux in the 
Lyman continuum ($\lambda \leq 912$~\AA) we determine a column density 
$N_{\rm HI} = (6\pm1) \times 10^{17}~{\rm cm}^{-2}$.  This measurement agrees with 
values (C.\ Gry \etal\ 2001) scaled from HST-observed column densities of \OI, but not 
with \NI.  The differences could arise from ionization effects or assumed metallicities
in the local ISM.  

\item Using non-LTE model atmospheres and observed EUV spectra, we estimate 
LL flux decrements $\Delta_{\rm star} = 110\pm10$ and $\Delta_{\rm ISM} = 45\pm5$, 
in agreement with the observed total decrement $\Delta_{\rm LL} = 5000\pm500$.  
The restored ionizing continuum of $\epsilon$~CMa outside the local clouds has 
photon flux $\Phi_{\rm LyC} \approx 3000\pm1000~{\rm cm}^{-2}~{\rm s}^{-1}$ and ionizing 
photon luminosity $Q_{\rm LyC} \approx 10^{45.7\pm0.3}~{\rm photons~s}^{-1}$ for 
$d = 124\pm2$~pc.  
 
\item  The \HI\ column density for the local cloud corresponds to optical depth 
$\tau_{\rm LL} = 3.8\pm0.1$ at the Lyman limit.   The EUV flux external to the local 
clouds is 10--20 times higher than viewed from Earth, with a photoionization rate 
$\Gamma_{\rm H} \approx$ (1--2)$\times 10^{-14}~{\rm s}^{-1}$.  At the cloud surface, 
hydrogen would be partially ionized, with ionization fraction $x \approx$ 30--40\% in
gas with $n_{\rm H} = 0.2~{\rm cm}^{-3}$ and $T \approx 7000$~K.  

\end{enumerate} 

\noindent
{\bf Future work.}   With an accurate interstellar \HI\ column density to $\epsilon$~CMa, the 
next step will be the development of 3D photoionization models of H, He, and heavy elements
in the local clouds.  As shown previously (J.\ Dupuis \etal\ 1995; J.\ Vallerga \& B.\ Welsh 1995) 
the EUV radiation field, as viewed from Earth, is dominated by five stellar sources:  two early 
B-type stars ($\epsilon$~CMa and $\beta$~CMa) and three bright white dwarf stars (G191-B2B, 
HZ~43, Feige~24).  While $\epsilon$~CMa is the dominant EUV source in the \HI\ ionizing band 
(504--912~\AA), the white dwarfs and EUV emission lines from surrounding hot gas dominate 
the EUV spectrum in the \HeI\ continuum (below 504~\AA). Recent analysis (R.\ Gladstone \etal\ 
2024) of the all-sky \Lya\ emission seen 
by the New Horizons spacecraft suggests that additional early B-type stars within the local hot 
bubble could contribute to the EUV radiation incident on the local clouds.   Ionizing photons from 
these sources enter the local clouds from different directions, requiring 3D models of the 
radiative transfer and geometric structure of 
the local clouds\footnote{C.\ Gry \& E.\ Jenkins (2014) suggested an alternative topology, in which 
the LIC and G-cloud are a single cloud, with internal velocity components arising from turbulence 
and a shock propagating through the cloud. S.\ Redfield \& J.\ Linsky (2015) evaluated this alternative 
using additional sight lines and concluded that multiple clouds provide better kinematic agreement.  
J.\ Linsky \& S.\ Redfield (2014) suggested cloud mixing along several sight lines along the axis 
between the LIC and G clouds.  In our models, we treat absorption from the clouds as a single 
source of flux attenuation.} .

\vspace{0.2cm}

\noindent
{\bf Time-dependent effects.}  
Most previous ionization models of the local ISM assumed equilibrium between photoionization
and radiative recombination.  However, over Myr timescales, these processes may not 
remain constant.  Within the ensemble known as the CLIC or ``complex of local interstellar clouds" 
(J.\ Slavin \& P.\ Frisch 2002), the Sun's motion of 25.7~\kms\ corresponds to 26.3~pc per Myr.  
Of the 15 clouds in the ISM within 15 pc (S.\ Redfield \& J.\ Linsky 2008) two dominate the angular 
coverage on the sky:  the local interstellar cloud (LIC) and the G cloud.  The Blue cloud and Aql cloud 
are additional components of the local ISM (J.\ Linsky \etal\ 2019).  P.\ Frisch (1994) suggested that 
the Sun entered the LIC within the last $10^4$~yr.   Subsequent papers (J.\ Linsky \etal\ 2019;  
J.\ Linsky \etal\ 2022) argued that the Sun and its heliosphere may exit the LIC within the next 2000 yrs.
P.\ Swaczyna \etal\ (2022) proposed that relative motions of LIC and G-cloud indicate a region of 
interaction.  

Another characteristic dynamical time scale comes from the motion of $\epsilon$~CMa 
relative to the Sun. 
\begin{equation}
    t_{\rm star} = d_*/V_* \approx (4.4~{\rm Myr}) \left( \frac {d_*}{124~{\rm pc}} \right)
                         \left( \frac {V_*}{27.3~{\rm km~s}^{-1}} \right)^{-1}  \;  \nonumber.
\end{equation}
From Hipparcos measurements (F.\ van Leeuwen 2007) its radial velocity is $V_* = 27.3\pm0.4$~\kms\ 
and its tangential velocity is $2.1\pm0.1$~\kms.  The latter velocity is based on its total proper motion 
$\mu_{\rm tot} = 3.50 \pm 0.14~{\rm mas~yr}^{-1}$ with components (in RA and Decl) of
 $\mu_{\alpha} \,\cos \delta  = 3.24 \pm 0.11~{\rm mas~yr}^{-1}$ and 
$\mu_{\delta} = 1.33 \pm 0.16~{\rm mas~yr}^{-1}$.  Approximately $4.4\pm0.1$~Myr ago, 
$\epsilon$~CMa passed by the Sun at an offset distance of $9.3\pm0.5$~pc, resulting in an
ionizing radiation field 100-200 times higher than its current value.   With the Sun's current velocity
setting it on a trajectory out of the local clouds, the location of $\epsilon$~CMa relative to the 
local hydrogen gas remains uncertain.  With a total mass of only 0.2--0.4~$M_{\odot}$, the local 
interstellar clouds within 3 pc are not self-gravitating and probably not gravitationally bound to the Sun. 
Thus, it is unclear whether they would be exposed to the same enhanced ionizing radiation.  
However, it is probable that $\epsilon$~CMa produced considerable effects in its wake.

We now estimate the characteristic time scales for hydrogen photoionization and recombination, 
and for radiative cooling of gas at the cloud surface:
\begin{eqnarray}
    t_{\rm ph}    &=& \Gamma_{\rm H}^{-1} \approx (1.6~{\rm Myr}) 
              \left( \frac {\Gamma_{\rm H}}{2\times10^{-14}~{\rm s}^{-1}}  \right)^{-1} \nonumber  \\
    t_{\rm rec}  &=& (n_e \alpha_{\rm H})^{-1}  \approx (0.93~{\rm Myr}) 
               \left( \frac {n_e}{0.1~{\rm cm}^{-3}} \right)^{-1} T_{7000}^{0.809} \nonumber  \\
    t_{\rm cool} &=&  \frac {3n_{\rm tot} kT/2} { n_{\rm H}^2 \, \Lambda(T)} 
              \approx (11~{\rm Myr}) \left( \frac {n_{\rm H}}{0.2~{\rm cm}^{-3}} \right)^{-1}  
                       T_{7000}   \nonumber  \; . 
\end{eqnarray}
Here, we scaled $\Gamma_{\rm H} $ to the estimated EUV continuum outside the LIC and adopted 
a case-B radiative recombination rate coefficient 
$\alpha_{\rm H} = (3.39\times10^{-13}~{\rm cm}^3~{\rm s}^{-1})T_{7000}^{-0.809}$ (B.\ Draine 2011),
appropriate for $T = (7000~{\rm K})T_{7000}$.  We adopted total hydrogen density 
$n_{\rm H} \approx 0.2$~cm$^{-3}$, electron density $n_e \approx 0.1~{\rm cm}^{-3}$, and a 
radiative cooling rate $n_{\rm H}^2 \Lambda(T)$ with coefficient 
$\Lambda(T) \approx 3\times10^{-26}~{\rm erg~cm}^3~{\rm s}^{-1}$ at 7000~K.   
The radiative cooling time and stellar crossing time are both longer than the time 
needed to establish photoionization equilibrium.

Approximately 4.4~Myr ago, $\epsilon$~CMa passed within 9--10~pc of the Sun, with a photoionization 
rate 180 times higher than at present.  Any gas clouds in the Sun's vicinity at that time would have 
been highly ionized.   In fact, $\epsilon$~CMa may have left a wake of ionized and photoelectrically 
heated gas, which could explain the tunnel of low N$_{\rm HI}$ in this direction.  Because 
$t_{\rm ph} \approx t_{\rm rec}$, an equilibrium level of ionization was established, evolving on Myr 
timescales tracking the B-star's motion.  In the future, the Sun will exit the local cloud and once again 
be exposed to a much higher ionizing radiation field unshielded by the local cloud.

\begin{acknowledgements}

 We thank the referee for comments on the spectral  classification and suggestions on 
 information displayed in figures.  
 We also thank Edward Jenkins for a thorough discussion of HST absorption-line studies
 of the sight line to $\epsilon$~CMa, including conversations at the Berkeley conference
 just two months before his decease.  His wisdom and contributions to studies of
 interstellar matter will be greatly missed.  We also thank James Green and Nick Erickson
 for access to their flux-calibrated EUV and FUV spectra of $\epsilon$~CMa, and 
 Cecile Gry, Jeffrey Linsky, Seth Redfield, and Jon Slavin for scientific discussions about the local 
 ISM.  John Hillier, Ivan Hubeny, and Tadziu Hoffmann kindly provided useful information 
 about their model atmosphere codes.  A portion of this study was supported by the 
 New Horizons Mission calibration observations of $\epsilon$~CMa and studies of cosmic 
 UV and \Lya\ backgrounds.

\end{acknowledgements}



 \clearpage



\begin{deluxetable} {lccc cccc}
\tablecolumns{8}
 \tabletypesize{\scriptsize}

\tablenum{1}
\tablewidth{0pt}
\tablecaption{Various Stellar Parameters\tablenotemark{a} }  

\tablehead{
   \colhead{Reference Paper}
 & \colhead{$d$}
 & \colhead{$T_{\rm eff}$} 
 & \colhead{$\log g$} 
 & \colhead{$R/R_{\odot}$}
 & \colhead{$M/M_{\odot}$} 
 & \colhead{$L/L_{\odot}$} 
 & \colhead{$M_{\rm bol}$}
 \\
 \colhead{}
 & \colhead{(pc)}
 & \colhead{(K)} 
 & \colhead{(cgs)} 
 & \colhead{}
 & \colhead{} 
 & \colhead{} 
 & \colhead{(mag)} 
 }

\startdata
Snow \& Morton (1976) & $\dots$ & 20,990 & 3.20 & 16.6 & 16 & 47,400 & $-6.95$ \\
Cassinelli \etal\ (1995)  & 188 & $20,990\pm760$ & $3.20\pm0.15$ & $16.2^{+1.2}_{-1.2}$  
                        & $15.2^{+6.4}_{-4.4}$ & $45,900\pm9500$ & $-6.91$  \\
Fossati \etal\ (2015)\tablenotemark{b} & $124\pm2$ & $22,500\pm300$ & $3.40\pm0.08$ & $12.0^{+1.7}_{-1.5}$ 
                         & $13.1^{+1.0}_{-0.9}$ & $33,250^{+9600}_{-8500}$ & $-6.56$ \\ 
Fossati \etal\ (2015)\tablenotemark{c}  & $124\pm2$ & $22,500\pm300$ & $3.40\pm0.08$ & $10.1^{+0.7}_{-0.5}$ 
                         & $12.0^{+0.4}_{-0.4}$ & $23,550^{+3500}_{-2650}$ & $-6.19$  \\ 
Erickson \etal\ (2021)   & $124\pm2$ & $22,500\pm300$ & $3.40\pm0.08$ & $12.0^{+1.7}_{-1.5}$ & 
                         $13.1^{+1.0}_{-0.9}$ & $33,250^{+9600}_{-8500}$ & $-6.56$  \\ 
Current Study (2024)   & $124\pm2$  & $21,000\pm780$ & $3.50\pm0.05$ & $10.7\pm0.7$ 
                         & $13.1\pm2.4$ & $19,900\pm1600$  & $-5.97$    \\
\enddata 

\tablenotetext{a} {Values of effective temperature, surface gravity, radius, mass, luminosity,
and bolometric absolute magnitude given in various papers.  L.\ Fossati \etal\ (2015) derived
$R$ and $M$ from two sets of evolutional tracks (see footnote b). }

\tablenotetext{b} {Stellar mass and radius inferred from evolutionary tracks of C.\ Georgy \etal\ (2013). }
\tablenotetext{c} {Stellar mass and radius inferred from evolutionary tracks of L.\ Brott \etal\ (2011). }

\end{deluxetable}



\begin{deluxetable} {lll}
\tablecolumns{3}
 \tabletypesize{\footnotesize}

\tablenum{2}
\tablewidth{0pt}
\tablecaption{Estimated \HI\ Column Densities\tablenotemark{a} }  

\tablehead{
   \colhead{Reference Paper}
 & \colhead{$N_{\rm HI}$ (cm$^{-2}$) }
 & \colhead{Method} 
 }

\startdata
J,\ Cassinelli \etal\ (1995)      &  (7--12)$\times10^{17}$        &  LyC (EUV-flux) modeling  \\
C.\ Gry \etal\ (1995)               &  $<5\times10^{17}$               &  N~I absorption (N/H = $6.76\times10^{-5}$)  \\
C.\ Gry \& E.\ Jenkins (2001) &  (3.1--3.7)$\times10^{17}$    &  N~I absorption (N/H = $7.5\times10^{-5}$) \\
C.\ Gry \& E.\ Jenkins (2001) &  (7--11)$\times10^{17}$        &  O~I  absorption (O/H = $3.16\times10^{-4}$)   \\
C.\ Gry \& E.\ Jenkins (2001) &  (4.8--7.2)$\times10^{17}$    &  O~I  absorption (O/H = $4.68\times10^{-4}$)   \\
Current Study                         & $(6\pm1)\times10^{17}$       &  LyC restoration + Model Atmospheres 
\enddata 

\tablenotetext{a} {Previous estimates of intervening column density of H~I.  
Methods include modeling the attenuation of stellar EUV fluxes and scaling 
$N_{\rm HI}$ from column densities of N~I or O~I, with assumptions about their
interstellar abundances and depletion factors.  The two values reported by
C.\ Gry \& E.\ Jenkins (2001) from O~I are based on different assumed (O/H)
abundances in the ISM ($3.16\times10^{-4}$) and in local B-stars 
($4.68\times10^{-4}$).  C.\ Gry \& E.\ Jenkins (2001) suggested that nitrogen 
may be underabundant relative to solar values.  There could also be variations 
in the ionization fractions of N, O, H, and He, owing to photoionization 
cross sections and charge exchange rates (O~I and N~I).  }

\end{deluxetable}



\begin{deluxetable} {llcccc}
\tablecolumns{6}
 \tabletypesize{\footnotesize}

\tablenum{3}
\tablewidth{0pt}
\tablecaption{Lyman Decrements and Photoionization Rates\tablenotemark{a} }  

\tablehead{
    \colhead{$N_{\rm HI}$  }
 & \colhead{$\Delta_{\rm LL}$ }
 & \colhead{$\Phi_{\rm LyC}$ } 
 & \colhead{$\Gamma_{\rm H}$ }  
 & \colhead{$x_{\rm s}$ }
 & \colhead{$x_{\rm s}$ }
  \\
    \colhead{ (cm$^{-2}$) }
 & \colhead{(ISM)}
 & \colhead{ (cm$^{-2}$~s$^{-1}$)} 
 & \colhead{ (s$^{-1})$ }  
 & \colhead{($0.2~{\rm cm}^{-3}$)}
 & \colhead{($0.1~{\rm cm}^{-3}$)}
 }

\startdata
 0 (no~abs)                 & 1.00  & 300    & $9.85\times10^{-16}$ & 0.108 & 0.150  \\
$3.0\times10^{17}$    &  6.62 & 883     & $3.35\times10^{-15}$ & 0.191 & 0.258  \\
$5.0\times10^{17}$    &  23.3 & 2000   & $8.25\times10^{-15}$ & 0.282 & 0.373  \\
$6.0\times10^{17}$    &  43.8 & 3090   & $1.33\times10^{-14}$ & 0.342 & 0.445  \\
$6.5\times10^{17}$    &  60.0 & 3870   & $1.69\times10^{-14}$ & 0.376 & 0.484  \\
$7.0\times10^{17}$    &  82.3 & 4870   & $2.17\times10^{-14}$ & 0.413 & 0.525  \\
$8.0\times10^{17}$    &  154  & 7810   & $3.58\times10^{-14}$ & 0.493 & 0.611  \\
$9.0\times10^{17}$    &  290  &12,700 & $6.00\times10^{-14}$ & 0.581 & 0.698    
\enddata 

\tablenotetext{a} {For the published range of interstellar column densities $N_{\rm HI}$, 
we present our modeled interstellar flux decrement $\Delta_{\rm LL} = \exp(\tau_{\rm LL})$ 
for optical depth $\tau_{\rm LL}$, integrated photon flux $\Phi_{\rm LyC}$ in the Lyman 
continuum, and photoionization rate  $\Gamma_{\rm H}$.  The last columns show 
the hydrogen ionization fraction, $x_{\rm s}$, at the cloud surface for two values,
$0.2~{\rm cm}^{-3}$ and $0.1~{\rm cm}^{-3}$, of total hydrogen density, 
$n_{\rm H} = n_{\rm HI} + n_{\rm HII}$.  We assume equilibrium 
$x^2/(1-x) = (\Gamma_{\rm H} / 1.1 n_{\rm H} \alpha_{\rm H})$ with recombination rate 
coefficient $\alpha_{\rm H} = 3.39 \times 10^{-13}~{\rm cm}^3~{\rm s}^{-1}$ at $T = 7000$~K. 
 }

\end{deluxetable}


\end{document}